\documentclass[10pt,conference]{IEEEtran}
\IEEEoverridecommandlockouts
\usepackage{cite}
\usepackage{amsmath,amssymb,amsfonts}
\usepackage{graphicx}
\usepackage{textcomp}
\usepackage{xcolor}
\def\BibTeX{{\rm B\kern-.05em{\sc i\kern-.025em b}\kern-.08em
    T\kern-.1667em\lower.7ex\hbox{E}\kern-.125emX}}
\usepackage[utf8]{inputenc}
\usepackage[english]{babel}
\usepackage[T1]{fontenc}
\usepackage{physics}
\usepackage{hyperref}
\usepackage{graphicx}
\usepackage{amsthm}

\newtheorem{theorem}{Theorem}
\newtheorem*{remark}{Remark}
\usepackage[font=itshape]{quoting}
\usepackage{xcolor}
\usepackage{qcircuit}
\usepackage{algpseudocode}
\usepackage{algorithm}
\usepackage{algorithmicx}
\usepackage{qcircuit}
\usepackage{subcaption}
\usepackage{amssymb}
\usepackage{orcidlink}

\newtheorem{proposition}{Proposition}
\newcommand{\Gate}[1]{\textsc{#1}}
\newcommand{\hgate}{\Gate{h}}
\newcommand{\zgate}{\Gate{z}}
\newcommand{\ygate}{\Gate{y}}
\newcommand{\xgate}{\Gate{x}}
\newcommand{\phasegate}{\Gate{s}}
\newcommand{\rzgate}{\Gate{Rz}}

\newcommand{\cnotgate}{\Gate{cnot}}

\begin{document}

\title{Quantum Error Mitigation by \\ Pauli Check Sandwiching
}

\author{\IEEEauthorblockN{Alvin Gonzales\orcidlink{0000-0003-1635-106X}\IEEEauthorrefmark{1}, Ruslan Shaydulin\orcidlink{0000-0002-8657-2848}\IEEEauthorrefmark{2}, Zain H. Saleem\IEEEauthorrefmark{3}, and Martin Suchara\IEEEauthorrefmark{3}\IEEEauthorrefmark{4}
\IEEEauthorblockA{\IEEEauthorrefmark{1}Intelligence Community Postdoctoral Research Fellowship Program, Argonne National Laboratory, Lemont, IL, USA}
\IEEEauthorblockA{\IEEEauthorrefmark{3}Mathematics and Computer Science Division, Argonne National Laboratory, Lemont, IL, USA}
\IEEEauthorblockA{\IEEEauthorrefmark{2}Global Technology Applied Research, JPMorgan Chase, New York, NY, USA}
\IEEEauthorblockA{\IEEEauthorrefmark{4}Amazon Web Services, Amazon, Seattle, WS, USA}
Email: \IEEEauthorrefmark{1}agonza@siu.edu}}


\maketitle


\maketitle


\begin{abstract}
    We describe and analyze an error mitigation technique that uses multiple pairs of parity checks to detect the presence of errors. Each pair of checks uses one ancilla qubit to detect a component of the error operator and represents one layer of the technique. We build on the results on extended flag gadgets and put it on a firm theoretical foundation. We prove that this technique can recover the noiseless state under the assumption of noise not affecting the checks. The method does not incur any encoding overhead and instead chooses the checks based on the input circuit. We provide an algorithm for obtaining such checks for an arbitrary target circuit. Since the method applies to any circuit and input state, it can be easily combined with other error mitigation techniques. We evaluate the performance of the proposed methods using extensive numerical simulations on 1,850 random input circuits composed of Clifford gates and non-Clifford single-qubit rotations, a class of circuits encompassing most commonly considered variational algorithm circuits. We observe average improvements in fidelity of 34 percentage points with six layers of checks. 
\end{abstract}

\section{Introduction}
Hardware errors or noise arising from qubit imperfections such as unwanted interactions with the environment limit the power of near-term quantum technologies. Since these devices lack the necessary number of qubits and error rates to perform quantum error correction \cite{Knill_1998FaultTolerQEC, aharonov1999_FaultTolerQEC, Kitaev1997_FaultTolerQEC}, error mitigation is required in order to increase the fidelity of computations. In this work we investigate error mitigation that uses a small number of ancillas to suppress the effect of errors. Various error mitigation techniques have been developed, such as zero-noise extrapolation ~\cite{Temme2017_ErrorMitigForShorDepthQuantCirc, 1805.04492, tudor2020_digitalZeroNoiseExtrapForQuantErrMitig}, which uses different error rates to reduce the error in the measurement of an observable; probabilistic error cancellation \cite{Temme2017_ErrorMitigForShorDepthQuantCirc}, which uses an ensemble of known noisy circuits to approach the correct expectation value; dynamical decoupling \cite{Viola_1998DynamicalDecoupling, Byrd_2003DetermineDynamicalDecouplingOps, Tripathi_2021SuppOfCrosstalkDynamicalDecoupling}, which uses timed control sequences to suppress interactions of the target quantum system with its environment; readout error mitigation \cite{Nachman_2020ReadoutNoise}, which uses classical postprocessing techniques to mitigate measurement errors; and symmetry verification \cite{Bonet-Monroig2018_LowCostErrMitigBySymmVerif, McArdle2019_ErrMitigDigitalQuantSimulation, shaydulin2021error, Cai2021quantumErrorMitigSymmExpan}, which verifies symmetries in computational problems of interest and discards erroneous computations.

Protocols that improve measurements of an observable have applications in problems such as the estimation of the ground state energy of a given Hamiltonian \cite{Peruzzo_2014AVariationalEignevalueSolver}. In contrast, protocols that improve fidelity generally apply to any problem. A main feature of many techniques aimed at improving measurements of an observable or reducing readout error 
is that they have no quantum overhead; in other words, they require no extra qubits or quantum operations (gates). Thus, these techniques are ideal for the noisy intermediate-scale quantum (NISQ) era \cite{Preskill2018quantumcomputingin} because current state-of-the-art NISQ devices contain few qubits, typically fewer than 50, and a limited number of gate operations because of fast decoherence times.

As quantum technology develops,  error mitigation schemes must adapt  and take advantage of improvements in qubit count and quality \cite{Gambetta_2021IBMRoadmap, chapman_2020IonqRoadmap}. Qubit count and error rates vary widely depending on the underlying qubit technology. Additionally, many of the current error mitigation techniques such as dynamical decoupling and probabilistic error cancellation require intricate tailoring of the protocol to the noise. Thus, they typically require the added overhead in costly quantum tomography \cite{Chuang_1997QuantumProcessTomo}.

In this work we theoretically and numerically study a quantum error mitigation technique inspired by stabilizer codes, that aims at improvement in quantum state fidelity. We build on the results of  \cite{Debroy_2020ExtendedFlagGadgets}, where they first explored the scheme of sandwiching a circuit between pairs of parity checks. Note that they refer to the pairs of checks as extended flag gadgets, inspired by the work in \cite{Chao2018_QECWithOnly2ExtraQubits, Zhou_2019QuantumCircsForDynamicRuntimeAssert}. Our research puts this parity check scheme on a firm theoretical foundation and numerically demonstrates its efficacy on a wide variety of quantum circuits. The main contributions of our work are: (1) extending the analysis to greater than two layers of checks, (2) establishing the theoretical limits of the technique, which culminates in the unit fidelity result of Theorem \ref{thm:MultilayerUnitFidelity}, (3) providing parity checks in Propositions \ref{lemma:weightOneErrors} and \ref{lemma:lowNumbOfChecks} that saturate this fidelity bound and hence answers an open question in \cite{Debroy_2020ExtendedFlagGadgets} regarding optimal checks to use, (4) providing a protocol that efficiently determines Pauli parity check pairs that can be used for a given input circuit, and (5) providing numerical simulations for a wide variety of random input circuits consisting of varying qubit count, \Gate{cnot} count, non-Clifford gate count, and layer count.

The error mitigation scheme that we study in this paper at its basic level of one layer uses one ancilla and two controlled unitary operations, which we refer to as checks. The parity checks sandwich the input circuit. Consequently, the error operator is conjugated between two controlled parity matrices. We measure the ancilla and postselect the state on the measurement outcomes. The net effect of the checks and the postselection is a transformed error map, where terms of the error map that anticommute with the checks are eliminated in the postselected state. The performance of this technique, measured by the improvement in quantum state fidelity, improves with the depth of the input circuit. Furthermore, this scheme is tunable, meaning that the number of layers and ancillary qubits used can be set by the user. 

This protocol shares some similarity to symmetry verification, which also uses stabilizer-style parity checks to improve the fidelity of the quantum state and requires no knowledge of the noise. However, unlike symmetry verification, which requires input states to be restricted to a specific eigenspace, this scheme places no restriction on the input state. Thus, the technique applies to subcircuits directly and can be easily combined with other error mitigation methods. 

In Theorem \ref{thm:MultilayerUnitFidelity}, we prove that in a restricted scenario where the noise does not affect the checks (see Figs. \ref{fig:multilayerNoisy} and \ref{fig:multilayerNoisyConjError}) there exist checks such that the postselected state is noiseless and the fidelity reaches unity. We provide an example of a randomly generated Clifford circuit with added checks that saturates this fidelity bound. 


We also investigate the performance of the scheme with numerical simulations in a more realistic setting where the checks are also noisy. The numerical simulations consist of $1,850$ (unmitigated) randomly generated five- and ten-qubit circuits composed of Clifford + arbitrary diagonal unitary gates. Our technique shows 
an average fidelity gain of 34 percentage points for random input circuits consisting of $40$ CNOTs with six layers of (noisy) checks;  see Figs. \ref{fig:5qubits_allCNOTs_Layers6_rz5}a and \ref{fig:5qubits_allCNOTs_Layers6_rz5}b. The increase in fidelity comes at a cost of a lower probability of postselecting on the ancillas' measurement outcomes. We also provide Clifford simulations that give intuition that this technique will perform well for deep circuits.


This paper is organized as follows. In Section \ref{sec:background} we review relevant background and provide definitions that are used in the paper. In Section \ref{subsec:SingleLayer} we provide the single-layer protocol. In Section \ref{subsec:postSelectedState} we describe the theoretical foundation of the technique. In Section \ref{subsec:MultilayerScheme} we provide the full multilayer scheme. In Section \ref{subsec:UpperBounds} we prove Theorem \ref{thm:MultilayerUnitFidelity} and  provide bounds on the number of layers required to reach unit fidelity  for the restricted scenario where the noise does not affect the checks. In Section \ref{subsec:generalErrors} we discuss how our results apply in general settings. In Section \ref{subsec:FindingChecks} we introduce techniques for finding checks quickly by using a precalculated table of commutation rules that eliminates the need to perform matrix multiplication. In Section \ref{subsec:NumericalResults} we give the results of our numerical simulations. In Section \ref{subsec:conclusions} we discuss our results and possible areas for future work.

\section{Background}\label{sec:background}


We begin with definitions and notation. For a detailed introduction to modeling of noise in quantum computation the reader is referred to Chapter 8 of~\cite{nielsen2011quantum}.

The most general evolution of an open quantum system is given by a dynamical map \cite{Sudarshan1961} 
\begin{align}
    \mathcal{E}:\rho_S\rightarrow\rho'_{S},
\end{align}
where $\rho_S$ and $\rho_S'$ are elements of the system Hilbert space $\mathcal{H}_S$. $\mathcal{H}_S$ is a subspace of the system and environment Hilbert space $\mathcal{H}_{SE}$, where $S$ is the system and $E$ is the environment. 

In the case of an initially unentangled system and environment, the map is completely positive and trace preserving. It can be derived by taking the partial trace of the global unitary evolution and yields the operator sum representation
\begin{align}\label{eq:cp_map}
    \notag\mathcal{E}(\rho_S)&=\tr_E(U_{SE}\rho_S\otimes\rho_EU_{SE}^\dagger)\\
    &=\sum_iE_i\rho_SE_i^\dagger,
\end{align}
where $U_{SE}$ is a unitary acting across the system and environment \cite{CHOI1975_CPMaps, Kraus_1983States}. The operators $E_i$ in Eq.~\eqref{eq:cp_map} are commonly called Kraus operators \cite{Kraus_1983States}. A map is completely positive (CP) if it maps all positive operators to positive operators when extended by the identity map to arbitrary higher dimensions~\cite{CHOI1975_CPMaps}, namely,
\begin{align}\label{eq:cp_condition}
\mathcal{E}\otimes\mathcal{I}_n(\rho)\geq 0 \quad\forall n, \rho,
\end{align}
where $\mathcal{I}_n$ is the $n$-dimensional identity map and $\rho$ is a density matrix. This extension to higher dimensions is required to ensure that when the input state is part of a higher-dimensional state, the output of the map is still positive. A map is trace preserving if
\begin{align}
    \sum_iE_i^\dagger E_i=\mathbb{I}.
\end{align}

Maps that do not satisfy Eq.~\eqref{eq:cp_condition} are called not completely positive (NCP) maps. NCP maps 
play a significant role in non-Markovian evolutions \cite{Milz_2019CPDivisibilityDoesNotMeanMarkov}, where the evolution of the state is often not decomposable into a sequence of completely positive maps. NCP maps have the form
\begin{align}
    \mathcal{E}(\rho)=\sum_i\eta_i E_i\rho E_i^\dagger,    
\end{align}
where $\eta_i=\pm 1$ and  at least one $\eta_i=-1$ exists \cite{CHOI1975_CPMaps}.

In this paper we use fidelity as a figure of merit. For two quantum states $\rho$ and $\omega$, fidelity is defined as
\begin{align}
    F(\rho, \omega)=\left(\tr\sqrt{\sqrt{\rho}\omega\sqrt{\rho}}\right)^2.
\end{align}
Fidelity is symmetric with regard to its inputs. 

Let $U$ denote the unitary operation implemented by the target circuit that we want to error mitigate. Let $\rho'$ be the density matrix representing the ideal, noiseless output of the target circuit, $\rho_n$ be the density matrix of the noisy quantum state produced by the target circuit, and $\rho_m$ be the density matrix of the noisy error-mitigated state produced by the error mitigated target circuit.

We denote the fidelity of the noisy state before error mitigation as $F_n=F(\rho_n, \rho')$, and the fidelity of the state after application of error mitigation as  $F_m=F(\rho_m, \rho')$.

We define the fidelity gain (improvement due to the technique) as
\begin{align}\label{eq:fidelityGain}
    F_m-F_n.
\end{align}
We say that the method ``detects all errors" when $F_m=1$ or equivalently when the error map $\mathcal{E}$ on the postselected state is identity, in other words, when all the Kraus operators of $\mathcal{E}$ are proportional to identity.

Next we describe the noise model used in our numerical simulations. The depolarizing channel for dimension $d$ is
\begin{align}
    \mathcal{D}_{p}(\rho)=(1-p)\rho+p\dfrac{\mathbb{I}}{d},
\end{align}
where $0\leq p\leq 1$ \cite{nielsen2011quantum}. For the numerical simulations of noisy circuits the only noiseless gates are a measurement gate or the input state, which is generated by a random circuit. Otherwise, we apply the single-qubit depolarizing channel after each single-qubit gate and the two-qubit depolarizing channel after every two-qubit gate. Throughout this paper, we set the two-qubit error rate to ten times the single-qubit error rate, an assumption that roughly corresponds to noise observed in current NISQ systems \cite{Patel_2020ExpEvalOfNISQQCErrMeasCharacAndImplic, Zhang2020_ErrorMitQuantGatesExceedPhysFidelities}.

\section{Methods}
In Sections \ref{subsec:SingleLayer} and \ref{subsec:postSelectedState} we describe the single-layer Pauli Check Sandwiching (PCS) technique and show that this protocol leads to a transformation of the error map. In Sections \ref{subsec:MultilayerScheme} and \ref{subsec:UpperBounds} we describe the multilayer protocol and prove that we can reach a fidelity of one between a noisy-mitigated circuit and a noiseless circuit when the error map is restricted to a subset of qubits. We also provide a small number of checks that achieve this fidelity. In Section \ref{subsec:generalErrors} we investigate how our techniques apply in a general setting. In Sections \ref{subsec:FindingChecks} and \ref{subsec:NumericalResults} we give the results of numerical experiments across $1,850$ unmitigated random circuits. 

\begin{figure}[!t]
    \centering
    \includegraphics[scale=1]{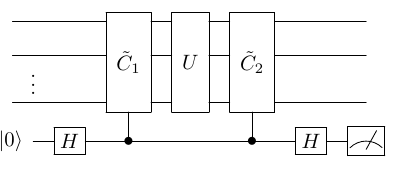}
    \caption{Overview of the one-layer version of the PCS scheme. $U$ represents the gates of the computation and acts across $n$ compute qubits. $U$ is sandwiched between two controlled unitaries comprising $\tilde{C}_1$ and $\tilde{C}_2$ that satisfy Eq.~\eqref{eq:conditionOnChecks1}.  The ancilla is the bottom qubit. The measurement is performed in the $\{\ket{0}, \ket{1}\}$ basis. The measurement outcome one is discarded, and zero is kept.}
    \label{fig:ourProtocolHighLevel}
\end{figure}
\subsection{Pauli Sandwich Error Mitigation Protocol: Single Layer}\label{subsec:SingleLayer}
We begin by describing the simplest version of the Pauli Check Sandwiching technique that consists of a single pair of parity checks sandwiching the computation (one ``layer''). Figure \ref{fig:ourProtocolHighLevel} shows a graphical view of the protocol. The unitary operation $U$ represents the gates of the computation. The bottom qubit is the single ancilla introduced by this scheme, and we commonly refer to the $n$ qubits above as the compute or computation qubits.

Let $C_2$ ($C_1$) be a controlled unitary with control on the ancilla that applies $\tilde{C}_2$ ($\tilde{C}_1$) on the compute target qubits. Mathematically, 
\begin{align}
    C_1&=\tilde{C}_1\otimes\op{1}+\mathbb{I}\otimes\op{0}\\
    C_2&=\tilde{C}_2\otimes\op{1}+\mathbb{I}\otimes\op{0}.
\end{align}
This scheme also requires that
\begin{align}\label{eq:conditionOnChecks1}
    \tilde{C}_2U\tilde{C}_1=U.
\end{align}

Before continuing, we make an important distinction between two protocols: the \textit{efficient PCS protocol} and the \textit{general PCS protocol}. For the \textit{efficient PCS protocol}, we restrict $\tilde{C}_1$ and $\tilde{C}_2$ to be elements of the $n$-qubit Pauli group $\mathcal{P}_n$, where 
\begin{align}
    \mathcal{P}_n=\{\mathbb{I},\xgate,\ygate, \zgate\}^{\otimes n}\times \{\pm 1, \pm i\}.
\end{align}
These added conditions are partly due to the difficult problem of determining the optimal circuit that implements $\tilde{C}_1$ from a given $\tilde{C}_2$ and $U$. Note that $\tilde{C}_2$ and $\tilde{C}_1$ can be much more general and still satisfy Eq.~\eqref{eq:conditionOnChecks1}. Thus, there are no additional constraints on the checks in the \textit{general PCS protocol}. 

In the \textit{general PCS protocol} and for a given $U$, any unitary $\tilde{C}_2$ can be used because in Eq.~\eqref{eq:conditionOnChecks1} we can always pick $\tilde{C}_2$ and solve for $\tilde{C}_1$.  We note in the text if a result holds for a specific case. If no statement is made, then the result holds for both scenarios. 

The single-layer protocol is as follows.
\begin{enumerate}
    \item Initialize the ancilla to $\ket{0}$ and apply a Hadamard gate. Perform $C_1$ with the control on the ancilla qubit and target on the compute qubits.
    \item Perform $U$ on the compute qubits.
    \item Perform $C_2$ with the control on the ancilla qubit and target on the compute qubits.
    \item Apply a Hadamard gate to the ancilla.
    Measure the ancilla in the $\{P_0=\op{0}, P_1=\op{1}\}$ basis, and discard the results where the outcome is $P_1$. We keep the result where the outcome is $P_0$.
\end{enumerate}


\subsection{Errors Detected by the Pauli Sandwich}
\label{subsec:postSelectedState}
\begin{figure}[!h]
    \begin{subfigure}{.5\textwidth}
        \centering
        \includegraphics[scale=1]{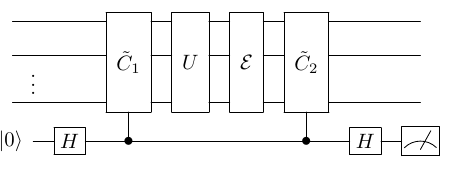}
        \caption{$\mathcal{E}$ is a noise map.}
        \label{fig:ourProtocolNoisy}
    \end{subfigure}
    \begin{subfigure}{.5\textwidth}
        \centering
        \includegraphics[scale=1]{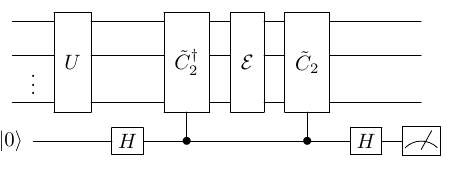}
        \caption{Equivalent noisy circuit using Eq.~\eqref{eq:conditionOnChecks1}. The gates after $U$ along with the postselection on the ancilla can be seen as the transformed error map.}
        \label{fig:singleLayerNoiseConjError}
    \end{subfigure}
    \caption{Noisy single-layer scheme}
\end{figure}
We now consider the effect of this scheme on an error map $\mathcal{E}$ acting on the compute qubits after $U$, as shown in Fig.~\ref{fig:ourProtocolNoisy}. Let $\mathcal{E}(\rho)=\sum_iE_i\rho E_i^\dagger$. Then the postselected output state of the protocol is
\begin{align}\label{eq:postSelectedState1}
    &\notag\rho_m=\\
    &\dfrac{\sum_i[(\tilde{C}_2E_i\tilde{C}_2^\dagger+E_i)U\rho U^\dagger (\tilde{C}_2E^\dagger_i\tilde{C}_2^\dagger+E^\dagger_i)]}{\tr(\sum_i[(\tilde{C}_2E_i\tilde{C}_2^\dagger+E_i)U\rho U^\dagger (\tilde{C}_2E^\dagger_i\tilde{C}_2^\dagger+E^\dagger_i)])}.
\end{align}
As shown in Fig.~\ref{fig:singleLayerNoiseConjError}, this protocol transforms the error map. We can write the postselected state given in Eq.~\eqref{eq:postSelectedState1} in terms of a new error map $\mathcal{E}'$,
\begin{align}\label{eq:postSelectedState2}
    \rho_m=\dfrac{\mathcal{E}'(U\rho U^\dagger)}{\tr[\mathcal{E}'(U\rho U^\dagger)]},
\end{align}
where $\mathcal{E}'$ has Kraus operators 
\begin{align}\label{eq:postSelectedError1}
    E'_i=\dfrac{\tilde{C}_2E_i\tilde{C}_2^\dagger+E_i}{2}
\end{align}
and the factor of $1/2$ comes from multiplying Eq.~\eqref{eq:postSelectedState1} by a convenient form of one, namely, $(1/4)/(1/4)$. 
We can now observe the power of this error mitigation technique. The error operators $E_i$ can be expanded in the Pauli basis. Thus, let
\begin{align}\label{eq:errorExpansion1}
    E_i=\sum_{\tilde\sigma_j\in \mathcal{P}_n}\alpha_{ij}\tilde\sigma_{j},
\end{align}
where $\tilde\sigma_{j}$ is an element of the Pauli group and $\alpha_{ij}=\tr(E_i\tilde\sigma_j)/(2^n)$ is a complex constant. Let $\tilde{C}_2\in\mathcal{P}_n$. Each $\tilde{\sigma}_j$ term in the expansion of $E_i$ either commutes or anticommutes with $\tilde{C}_2$. Substituting Eq.~\eqref{eq:errorExpansion1} into Eq.~\eqref{eq:postSelectedError1}, we see that the $\tilde{\sigma}_{j}$ elements that anticommute with $\tilde{C}_2$ are eliminated and
\begin{align}\label{eq:postSelectedError2}
    E'_i=\sum_{\tilde\sigma_j\in \mathcal{P}'_n}\alpha_{ij}\tilde\sigma_{j},
\end{align}
where $\mathcal{P}'_n$ is the Pauli group excluding the elements that anticommute with $\tilde{C}_2$. 

The effect of the protocol on the error map shares some similarities to that of twirling \cite{Cai2019_ConstructingSmallerPauliTwirlSetsForArbitErrChannels, Bennett1996_MixedStateEntAndQEC, Bennett1996_PurificationOfNoisyEntAndFaithfulTelepViaNoisyChannels}. In twirling, the twirling set $T$ is used to conjugate the error map:
\begin{align}
    \notag&\dfrac{1}{\abs{T}}\sum_{V\in T}V\mathcal{E}(V^\dagger\rho V)V^\dagger\\
    &=\dfrac{1}{\abs{T}}\sum_{i,(V\in T)}VE_i V^\dagger\rho VE_i^\dagger V^\dagger.
\end{align}
Usually, twirling is performed by using the Pauli or Clifford group as the twirling set. When twirling is performed with a suitable set, it transforms the noise into a Pauli channel. However, the PCS scheme is in some sense more powerful since it completely eliminates the contribution of anticommuting Pauli terms. 

\begin{figure}[!h]
    \centering
    \includegraphics[scale=.6]{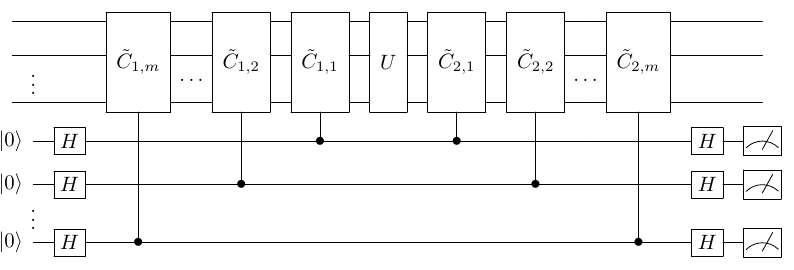}
    \caption{Multilayer scheme. There are $n$ compute qubits, $m$ layers, and $m$ ancillas. The second index in the controlled unitaries represents the layer. Each layer uses one ancilla and two checks. The checks sandwich the input circuit.}
    \label{fig:multilayerNoiseless}
\end{figure}

\subsection{Pauli Sandwich Error Mitigation Protocol: Multiple Layers}\label{subsec:MultilayerScheme}

The suppression of errors from anticommuting Pauli terms in the postselected state can be enhanced by introducing multiple layers of the single-layer error mitigation technique. A graphical view of how this scheme works is given in Fig.~\ref{fig:multilayerNoiseless}. There are $m$ layers with each layer consisting of controlled operations $C_{1,k}$ and $ C_{2,k}$, where the second index represents the layer, and one ancilla corresponding to each layer. Each pair of $C_{1,k}$ and $C_{2,k}$ satisfies
\begin{align}\label{eq:multiLayerConditionOnChecks1}
    \tilde{C}_{2,k}U\tilde{C}_{1,k}=U.
\end{align}
The multilayer scheme generalizes the single-layer scheme and is performed as follows.
\begin{enumerate}
    \item Initialize the ancillas to $\ket{0}$, and perform Hadamard gates on the ancillas. Perform $C_{1,k}$ with control on the $k^\text{th}$ ancilla qubit and target on the compute qubits.
    \item Perform $U$ on the compute qubits.
    \item Perform $C_{2,k}$ with control on the $k^\text{th}$ ancilla qubit and target on the compute qubits.
    \item Perform Hadamard gates on the ancillas. Measure all the ancillas in the $\{P_0=\op{0}, P_1=\op{1}\}$ basis, and discard the results where at least one of the outcomes is $P_1$. We keep the result where all the outcomes are $P_0$.
\end{enumerate}

\subsection{Upper Bounds on Fidelity and Required Number of Checks}\label{subsec:UpperBounds}
\begin{figure}[!h]
    \begin{subfigure}[b]{.5\textwidth}
        \centering
        \includegraphics[scale=.6]{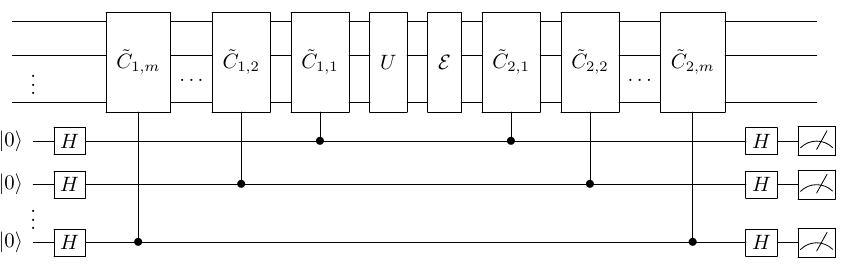}
        \caption{$\mathcal{E}$ is an arbitrary noise map.}
        \label{fig:multilayerNoisy}
    \end{subfigure}
    \begin{subfigure}[b]{.5\textwidth}
        \centering
        \includegraphics[scale=.6]{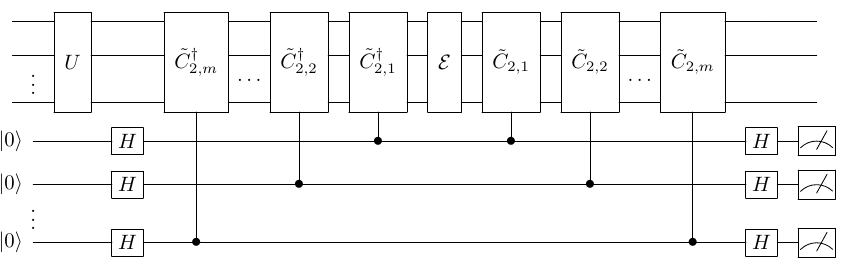}
        \caption{Equivalent circuit using  Eq.~\eqref{eq:multiLayerConditionOnChecks1}. Reminiscent of the single-layer scheme, the gates after $U$ along with the postselection on the ancillas can be seen as the transformed error map.}
        \label{fig:multilayerNoisyConjError}
    \end{subfigure}
    \caption{Noisy multilayer scheme}
\end{figure}


Now let us consider a noise map $\mathcal{E}(\rho)=\sum_i E_i\rho E_i^\dagger$ acting after $U$ on a subset of qubits as shown in Fig.~\ref{fig:multilayerNoisy}. From Eq.~\eqref{eq:postSelectedError1}, in the expansion of $E_i$ in the Pauli basis, we know that the $k$th layer eliminates Pauli terms that anticommute with $\tilde{C}_{2,k}$. This immediately leads to the observation that we can detect all errors under the noise model shown in Fig.~\ref{fig:multilayerNoisy}, which we prove in the following theorem. Theorem \ref{thm:MultilayerUnitFidelity} holds in general for the \textit{general PCS protocol} and it holds for the \textit{efficient PCS protocol} when the checks are in the Pauli group, in other words, when $U$ is Clifford. Note that we discuss why these results hold for NCP errors as well later at the start of Section \ref{subsec:generalErrors}.
\begin{theorem}[Unit Fidelity]\label{thm:MultilayerUnitFidelity}
    If errors are restricted to act only on the compute qubits, for any noisy unitary quantum circuit $U$ acting on $n$ compute qubits, there exist checks (see Fig.~\ref{fig:multilayerNoisy}) such that the fidelity between the post selected state and a noiseless run (noiseless execution of U only) reaches one.
\end{theorem}
\begin{proof}
    First, note that if the error map $\mathcal{E}(\rho)=\sum_iE_i\rho E_i^\dagger$ is the identity map, then the fidelity between the output $\rho_m$ of the error-mitigated circuit and the output $\rho'$ of a circuit with only $U$ (a noiseless run) is
    \begin{align}
        F(\rho_m,\rho')=1.
    \end{align}
    This directly follows from Eq.~\eqref{eq:multiLayerConditionOnChecks1}. Thus, if we can transform all  the Kraus operators $E_i$ of the error $\mathcal{E}$ to identity in the postselected state, then we have the result.
    
    Notice that from Eq.~\eqref{eq:multiLayerConditionOnChecks1}, Fig.~\ref{fig:multilayerNoisy} is equivalent to Fig.~\ref{fig:multilayerNoisyConjError} and the error map is conjugated by multiple layers of checks. Expanding the error in the Pauli basis, we have
    \begin{align}
        E_i=\sum_{\tilde\sigma_j\in \mathcal{P}_n}\alpha_{ij}\tilde\sigma_{j},
    \end{align}
    where $\tilde\sigma_{j}$ is an element of the Pauli group $\mathcal{P}_n$ and $\alpha_{ij}=\tr(E_i\tilde\sigma_j)/(2^n)$ is a complex constant. Let $\tilde{C}_{2,i}\in \mathcal{P}_n, \forall i$.
    
    We now make the results given in Section \ref{subsec:postSelectedState} recursive. First, we label the check layers from 1 to $m$ starting with the innermost layer. Then, Eq.~\eqref{eq:postSelectedError1} can be written recursively as
    \begin{align}\label{eq:multilayerPostError1}
        E_i^{(k)}=\dfrac{\tilde{C}_{2,k}E_i^{(k-1)}\tilde{C}_{2,k}^\dagger+E_i^{(k-1)}}{2},
    \end{align}
    where $(k)$ represents the layer and $E_i^{(0)}$ is the initial error Kraus operation. This leads to the recursive form of Eq.~\eqref{eq:postSelectedError2},
    \begin{align}\label{eq:multilayerPostError2}
         E^{(k)}_i=\sum_{\sigma_j\in G^{(k)}_n}\alpha_{ij}\tilde\sigma_{j},
    \end{align}
    where $G^{(k)}_n$ is the Pauli group excluding the elements that anticommute with $\{\tilde{C}_{2,1}, \tilde{C}_{2,2}, \cdots, \tilde{C}_{2,k}\}$. 
    Letting $k$ equal the size of $\mathcal{P}_n$ (excluding global phases), namely, $4^n$, we get $E_i^{(k)}=\alpha_i I$. The $\alpha_i$ is a constant that cancels out under renormalization, and the result follows.
\end{proof}


Before proceeding, we need to clarify the implications of Theorem \ref{thm:MultilayerUnitFidelity}. In 
that theorem, if we satisfy the conditions, we will have unit fidelity in the postselected state. However, the probability of postselecting is 
\begin{align}
    P(\overline 0)=\tr(\mathcal{E}^{(m)}(U\rho U^\dagger)),
\end{align}
where the Kraus operators of $\mathcal{E}^{(m)}$ are given by Eq.~\eqref{eq:multilayerPostError2}.
In Eq.~\eqref{eq:multilayerPostError2} the checks will eliminate all  the Pauli terms that are nonidentity. Thus, we see that if all  the Kraus operators of the error map are traceless, in other words, contain no identity term in their expansion in the Pauli basis, all  the Kraus operators for the error map in the postselected state will be the zero matrix, and the probability of postselecting is zero. This makes sense because we are not correcting errors, but mitigating errors by post selecting outcomes. The theorem holds trivially in this scenario because there is no post selected state.

Moreover, from Fig.~\ref{fig:multilayerNoisy} and Theorem \ref{thm:MultilayerUnitFidelity}, it seems that we can set $\mathcal{E}(\rho)=U^\dagger \rho U$, which eliminates $U$, and  use only the checks for the implementation of the circuit. While this is certainly true, we must consider the postselection probability. If $U$ is traceless, the probability of postselecting is zero.

Next we provide a small number of $C_2$ checks that can reach unit fidelity in the setting of Theorem \ref{thm:MultilayerUnitFidelity}. The following results given in Propositions \ref{lemma:weightOneErrors} and \ref{lemma:lowNumbOfChecks} are for the \textit{general PCS protocol}. Propositions \ref{lemma:weightOneErrors} and \ref{lemma:lowNumbOfChecks} hold for the \textit{efficient PCS protocol} given that the $\tilde{C}_1$ checks are in the Pauli group, in other words, $U$ is Clifford. Propositions \ref{lemma:weightOneErrors} and \ref{lemma:lowNumbOfChecks} hold for the noise model give in Fig.~\ref{fig:multilayerNoisy}. For arbitrary weight-one Kraus errors, that is, the Kraus error operators, $E_i$ act only on a single qubit; there exist two layers, where $\tilde{C}_{2,1}$ and $\tilde{C}_{2,2}$ are max weight, and we reach unit fidelity in the postselected state.
\begin{proposition}[Weight-One Kraus Operators: Two layers of max weight checks are sufficient]\label{lemma:weightOneErrors}
    For the noise model given in Fig.~\ref{fig:multilayerNoisy} and for all $\mathcal{E}$ consisting of only weight-one $E_i$, there exist two layers of checks such that we have unit fidelity in the postselected state. The $C_2$ part of the checks requires a total of $2n$ $\Gate{cnot}$ gates, where $n$ is the number of compute qubits.
\end{proposition}
\begin{proof}
    Each of the single-qubit errors can be expanded in terms of the single-qubit Pauli gates. Thus,
    \begin{align}\label{eq:weight1Expansion1}
        E_{i,k}=\sum_j\alpha_{i,j}\sigma_{j,k},
    \end{align}
    where $k$ is the qubit it is acting on, $\sigma_j$ is a Pauli matrix or identity, and $\alpha_{i,j}$ is a complex constant. Let our checks be
    \begin{align}
        \tilde{C}_{2,1}=\xgate^{\otimes{n}}
    \end{align}
    and
    \begin{align}
        \tilde{C}_{2,2}=\zgate^{\otimes{n}}.
    \end{align}
    These checks are inspired by the parity checks used in Shor's code \cite{Shor_1995QEC}.
    The $\tilde{C}_{2,1}$ consist of tensors of Pauli X and anticommutes with Pauli Y and Pauli Z errors in Eq.~\eqref{eq:weight1Expansion1}. The $\tilde{C}_{2,2}$ consist of tensors of Pauli Z and anticommutes with Pauli X errors in Eq.~\eqref{eq:weight1Expansion1}. From Theorem \ref{thm:MultilayerUnitFidelity}, the anticommuting terms in the error operators are suppressed. Thus, these two layers of checks are sufficient to reach fidelity one.
\end{proof}
The checks given in Proposition \ref{lemma:weightOneErrors} can detect all errors $\mathcal{E}$ that consist of weight-one Kraus operators $E_i$. This class of errors contains error maps that are more general than just single-qubit error maps. For example, $E_1$ can act on qubit one, and $E_2$ can act on qubit two. $E_1$ and $E_2$ are weight-one errors, but the overall map affects multiple qubits. 
\begin{remark}
    At least two layers are necessary to reach fidelity one in Proposition \ref{lemma:weightOneErrors}. To see this, we need only  show that a single layer is insufficient for arbitrary weight-one errors. Consider a circuit with only one compute qubit. For an arbitrary single-layer scheme, let $\tilde{C}_2 = W$ be the check. Then let the error map be
    $E=W$. The check and the error do not anticommute so the error map in the postselected state is not identity. Thus, a single layer is insufficient to detect all weight-one errors;  at least two layers are necessary. Proposition \ref{lemma:weightOneErrors} shows that we can always saturate this lower bound on the number of required checks.
\end{remark}

We can also reach fidelity one for arbitrary weight errors for the error model given in Fig.~\ref{fig:multilayerNoisy} with a small number of weight-one $\tilde{C}_{2,k}$. These checks are generators of the Pauli group and require $2n$ layers, but the $C_2$ components of the checks require the same number of $\Gate{cnot}$ gates as in Proposition \ref{lemma:weightOneErrors}. Thus, generally at the cost of more ancillas, we can detect all errors on the postselected state. Consider two weight-one $\tilde{C}_2$ checks of $\sigma_1$ and $\sigma_3$ on the $k$th compute qubit. All Pauli group elements that are nonidentity on the $k$th qubit anticommute with either $\sigma_1$ or $\sigma_3$. This leads to the following small set that can reach fidelity one.
\begin{proposition}[Any Error: $2n$ number of weight-one checks are sufficient]\label{lemma:lowNumbOfChecks}
    For the noise model given in Fig.~\ref{fig:multilayerNoisy} and arbitrary errors, let $n$ be the number of compute qubits. Then there exist $2n$ number of distinct (ignoring the global phase) weight-one $\tilde{C}_{2, k}$ such that we have unit fidelity in the postselected state.
\end{proposition}
\begin{proof}
    Let the $k$th compute qubit have two layers acting on it with $\tilde{C}^{(k)}_{2,r}=\sigma_1^{(k)}$ and $\tilde{C}^{(k)}_{2,l}=\sigma_3^{(k)}$. All Pauli group elements that are nonidentity on the $k$th qubit anticommute with at least one of the checks. Thus, this eliminates all Pauli terms in the expansion of the error Kraus operators that do not have identity on the $k$th qubit for the postselected state. We repeat these checks for the other compute qubits. The same argument holds in general for $\{\sigma_i^{(k)}, \sigma_j^{(k)} |i\neq j\}$ and the result follows.
\end{proof}

\begin{figure}[!h]
    \centering
    \includegraphics[scale=.5]{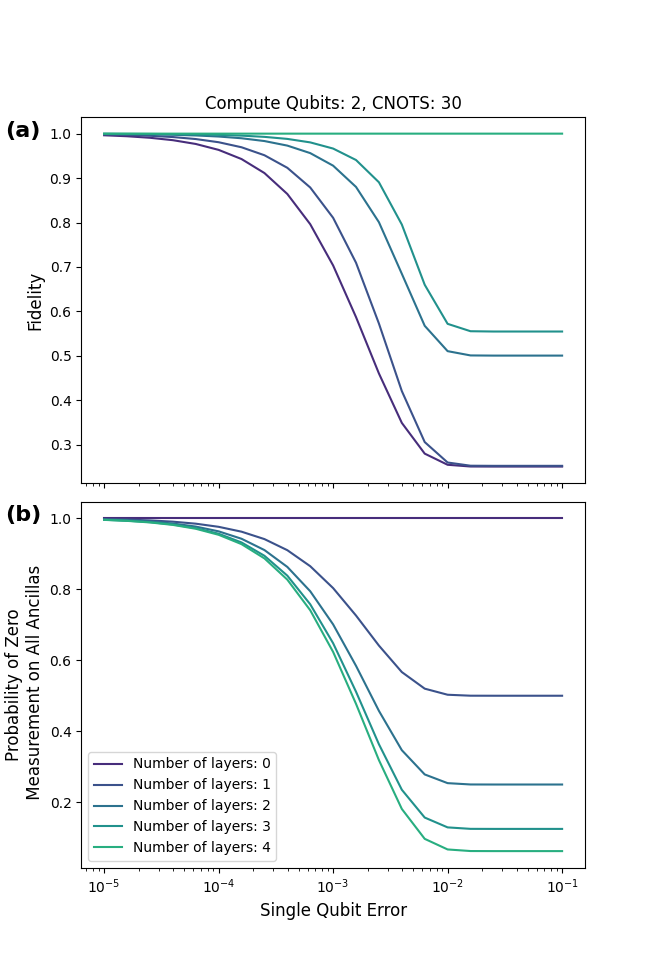}
    \caption{Example of checks that detect all errors. The upper bound on fidelity is saturated at four layers for this randomly generated Clifford input circuit consisting of two qubits and 30 \Gate{cnot} gates. We use depolarizing noise for the given noise model in Fig.~\ref{fig:multilayerNoisy}. The two-qubit error rate is ten times the single-qubit error rate. The single qubit-error rate ranges from $10^{-5}$ to $10^{-1}$. At $10^{-1}$, each \Gate{cnot} gate (acting on the compute qubits only) is followed by a two-qubit maximal depolarizing channel. Regardless, the postselected state is noiseless, as predicted by Theorem \ref{thm:MultilayerUnitFidelity}.}
    \label{fig:2qubit_saturate_fidelity}
\end{figure}
Figure \ref{fig:2qubit_saturate_fidelity} shows an example of a random Clifford circuit consisting of two compute qubits and 30 \Gate{cnot} gates that gives unit fidelity for the postselected state. This matches the prediction of Theorem \ref{thm:MultilayerUnitFidelity}. We use a Clifford circuit to guarantee that we can get the desired checks with the \textit{efficient PCS protocol}. We use the checks provided in Proposition \ref{lemma:lowNumbOfChecks}. The two checks on each compute qubit are $\sigma_1^{(k)}$ and $\sigma_2^{(k)}$, and we vary the number of layers from zero to four. Interestingly, at the single-qubit error of $0.1$, the two-qubit depolarizing channel is maximally depolarizing, but the fidelity remains at one for the postselected state (as predicted).

The gain in fidelity comes at the cost of a lower probability of measuring all zeros for the ancillas. This trade-off is demonstrated in Fig.~\ref{fig:2qubit_saturate_fidelity}b. The probability of measuring all zeros $p(\overline 0)$ drops to around 7\% for this circuit at the high single-qubit error of $0.1$. Note that the overhead in the number of runs is $\frac{1}{p(\overline 0)}$.

\subsection{General Errors and Hardware Considerations}\label{subsec:generalErrors}
\begin{figure}[!h]
    \begin{subfigure}[b]{.5\textwidth}
        \centering
        \includegraphics[scale=.6]{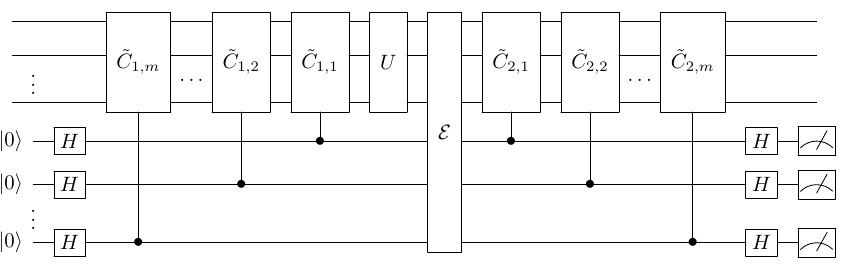}
        \caption{$\mathcal{E}$ is a general noise map that acts across all qubits.}
        \label{fig:multilayerNoisyGeneral}
    \end{subfigure}
    \begin{subfigure}[b]{.5\textwidth}
        \centering
        \includegraphics[scale=.6]{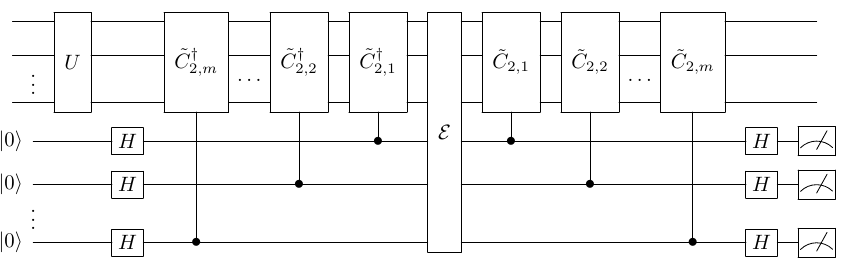}
        \caption{Using Eq.~\eqref{eq:multiLayerConditionOnChecks1}, the error map is still conjugated between the checks.}
        \label{fig:multilayerNoisyGeneralConjError}
    \end{subfigure}
    \caption{Noisy multilayer scheme}
\end{figure}
In the preceding sections,  we restricted $\mathcal{E}$ to CP maps, but our results hold also for general linear Hermitian maps, which includes NCP maps. As previously mentioned, NCP maps play a major role in non-Markovian evolutions, where the maps tend to be non-CP divisible. NCP maps have a  similar form to CP maps and are written as $\mathcal{E}(\rho)=\sum_i\eta_i E_i\rho E_i^\dagger$, where $\eta_i=\pm 1$ and there exists at least one $\eta_i=-1$. The coefficients $\eta_i$ are not used in any of our proofs, and thus the results hold.

Also, we restricted $\mathcal{E}$ to act only on the compute qubits. Obviously this is a restricted case, and in physical systems the checks are noisy and the error map would generally act across all the qubits, as shown in Fig.~\ref{fig:multilayerNoisyGeneral}. In this situation, the checks still conjugate the error, as shown in Fig.~\ref{fig:multilayerNoisyGeneralConjError}. Consequently, the technique is effective when $\mathcal{E}$ is dominated by Kraus operators that mainly affect the compute qubits; that is, the majority of the noise is from $U$. 

On non fully connected quantum computers, the parity checks may be difficult to perform with resulting minimal noise on the ancillas due to the need for swapping qubits. Thus, applications of this technique likely need to carefully map the circuit to the hardware to minimize the swaps between ancillas and compute qubits or execute the circuits on a fully connected device. 

Since single-qubit gates introduce less noise than nonlocal gates, the Pauli group is a good candidate for the $\tilde C_2$ part of the checks.  Furthermore, when $U$ is a deep circuit, the noise it induces will generally act across multiple qubits. In this scenario, low-weight $\tilde C_2$ will act nontrivially on these errors.  Thus, in general it is better to use low-weight checks in order to avoid introducing too many errors.

Moreover, for some executions of this scheme, the postselection probability may be smaller than desired. The postselection probability can be increased by reducing the number of check layers.

\subsection{Protocol for Finding Checks Quickly}\label{subsec:FindingChecks}

\begin{figure}
    \begin{subfigure}[b]{.5\textwidth}
        \centering
        \includegraphics[scale=1]{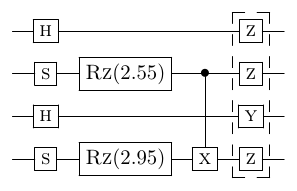}
        \caption{We assign $\zgate\ygate\zgate\zgate$ to $\tilde C_2$}
    \end{subfigure}
    \begin{subfigure}[b]{0.5\textwidth}
        \centering
        \includegraphics[scale=1]{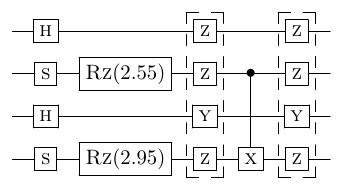}
        \caption{We propagate one layer to the left and determine the intermediate check gate $\zgate\ygate\zgate\zgate$.}
    \end{subfigure}
    \begin{subfigure}[b]{0.5\textwidth}
        \centering
        \includegraphics[scale=1]{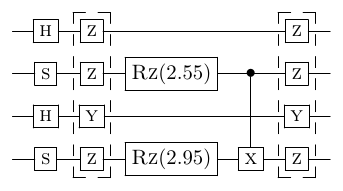}
        \caption{We propagate the intermediate check gate one layer to the left and determine the intermediate check gate $\zgate\ygate\zgate\zgate$. Notice that the $\zgate$ gate commutes with $\rzgate$.}
    \end{subfigure}
    \begin{subfigure}[b]{0.5\textwidth}
        \centering
        \includegraphics[scale=1]{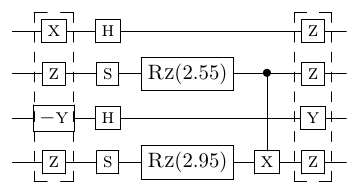}
        \caption{We propagate through the last layer and  assign the result $-\zgate\ygate\zgate\xgate$ to the $\tilde C_1$ gate.}
    \end{subfigure}
    \caption{Visual example of ``pushing" the checks through the circuit. We start with $\tilde C_2$ and find $\tilde C_1$. No multiplication is performed since the propagation of the $\tilde C_2$ gate is determined through lookups of predetermined commutation relations.}
    \label{fig:checksExample}
\end{figure}
\begin{figure}[!h]
        \centering
        \includegraphics[scale=.75]{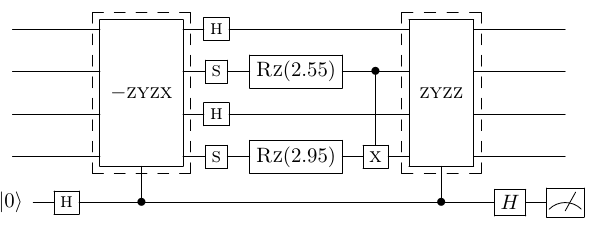}
        \caption{Final error-mitigated circuit for the example described in Fig.~\ref{fig:checksExample}}
        \label{fig:completedChecksExample}
\end{figure}
While checks always exist for a given $U$, in practice it is difficult to directly compute $\tilde{C}_1$ from Eq.~\eqref{eq:multiLayerConditionOnChecks1} for a given $\tilde{C}_2$. Here we introduce our searching protocol for the \textit{efficient PCS protocol} for determining the check pairs quickly and without matrix multiplication. Note that this protocol can fail to find any checks or may not find the desired number of checks. This can happen when the circuit contains many non-Clifford gates. We refer to the checks searching protocol as the \textit{finding checks protocol}. For our implementation, we constrained $\tilde{C}_1$ and $\tilde{C}_2$ to be in the Pauli group. We leave the potential searching protocol of a non-Pauli $\tilde{C}_1$ for future work.

The goal is to determine the gates comprising $\tilde C_1$ from a given $\tilde C_2\in\mathcal{P}_n$ and a given $U$. Instead of performing matrix multiplication, we transpile the input circuit to an equivalent circuit that uses the gate set $\{\xgate, \ygate, \rzgate, \phasegate, \hgate, \cnotgate\}$ and perform lookups of the commutation relations. This method applies to circuits consisting of Clifford $+$ arbitrary diagonal gates, which is a universal gate set since diagonal gates contain the gate $T$. 

To determine the checks, we use the equality $U_1 U_2= U_2 (U_2^\dagger U_1 U_2)=U_2 U'_1$, where $U'_1=U_2^\dagger U_1 U_2$ and $U_1$ and $U_2$ are unitary. We refer to this technique as ``pushing" $U_1$ through $U_2$. Figure \ref{fig:checksExample} gives a visual example of the pushing of the $\tilde{C}_2$ gates to determine $\tilde C_1$. Figure \ref{fig:completedChecksExample} is the completed error-mitigated circuit. This process is efficient since the cost of each lookup call is constant $O(1)$. 

Algorithm \ref{alg:main} is the pseudocode for the main script for finding a desired number of Pauli check pairs. It iterates through the minimum weight Pauli checks first and terminates when a sufficient number of layers of checks have been found. The protocol focuses on using low weight checks to minimize the noise introduced by the checks as discussed previously in Section \ref{subsec:generalErrors}. The main script calls on Alg. \ref{alg:canContinue} to see whether it is possible to push the current gate through. In Alg. \ref{alg:pushGateThrough} the lookup call is a preset table that has commutation relations. This symbolic ``pushing" of Pauli gates through $U$ works for all gates in the basis set except for \Gate{Rz}. For $\Gate{Rz}$, if the gate being pushed is not in $\{\Gate{z}, \Gate{i}\}$, which are operators that commute with an arbitrary diagonal gate, then we skip that Pauli group element.

Note that mathematically, any $\tilde{C}_2\in \mathcal{P}_n$ can be used because $\tilde{C}_1$ can be determined from Eq.~\eqref{eq:conditionOnChecks1}. Thus, one should be able to expand the current algorithm to allow for finding of general $\tilde{C}_1$. This problem is nontrivial.
\begin{algorithm}
\caption{Main script: find pairs of Pauli checks}
\label{alg:main}
\begin{algorithmic}[1]
    \State $circ \gets$ quantum circuit
    \State $paulis \gets$ +1 phase Pauli group for N qubits.
    $paulis$ is sorted by weight from smallest to largest
    \State $c_1$ is initialized to None
    \State $c_2$ is initialized to None
    \State $layersFound \gets 0$
    \State $numberLayers \gets$ number of layers to find for $circ$.
    \For{$pauli$ in $paulis$}
        \If{$layersFound$ equals $numberLayers$}
            \State BREAK \Comment{Found the necessary number of layers.}
        \EndIf
        \State $op2\gets pauli$
        \For{$op1$ and $index$ in $circ$}
            \If{\Call{CanContinue}{$op1$, $op2$}}
                \State $op2 \gets$ \Call{Push}{op1, $op2$}
            \Else
                \State BREAK \Comment{This check attempt didn't work so break out and try the next Pauli.}
            \EndIf
            \If{$index$ is the last index}
                \State $layersFound \gets layersFound+1$
                \State $c_1 \gets$ append $op2$
                \State $c_2 \gets$ append $pauli$ 
            \EndIf
        \EndFor
    \EndFor
\end{algorithmic}
\end{algorithm}

\begin{algorithm}
\caption{Push Pauli gate left}
\label{alg:pushGateThrough}
\begin{algorithmic}[1]
    \State \textbf{Input:}\\
    op1: The current gate that we need to pass through.\\
    \indent op2: The gate being pushed through gate.
    \State \textbf{Output:}\\
    The gate after it is pushed through op1. Should be equivalent to $op1^\dagger*op2*op1$
    \Function{push}{$op1$, $op2$}
        \If{$op1$ is X}
            \State \Return lookup $X(op2)X$
        \ElsIf{$op1$ is Y}
            \State \Return lookup $Y(op2)Y$
        \ElsIf{$op1$ is H}
            \State \Return lookup $H(op2)H$
        \ElsIf{$op1$ is S}
            \State \Return lookup $S^\dagger(op2)S$
        \ElsIf{$op1$ is \Gate{Rz}}
            \State \Return $op1$ \Comment{$op1$ will commute with \Gate{Rz} from the previous check performed in the function CanContinue}
        \ElsIf{$op1$ is \Gate{cnot}}
            \State \Return lookup $CX(op2)CX$
        \EndIf
    \EndFunction
\end{algorithmic}
\end{algorithm}

\begin{algorithm}
\caption{Checks if it is possible to pass the gate through}
\label{alg:canContinue}
\begin{algorithmic}[1]
    \State \textbf{Input:}\\
    op1: The current gate that we need to pass through.\\
    \indent op2: The gate being pushed through gate.
    \State \textbf{Output:}\\
    True if the gate can be pushed through and false if not.
    \Function{CanContinue}{$op1$, $op2$}
        \State\Return $op1$ is not $\Gate{Rz}$ or ($op2$ is I or Z)
    \EndFunction
\end{algorithmic}
\end{algorithm}

\subsection{Numerical Results}\label{subsec:NumericalResults}
The analytical results presented above assume perfect checks. Here we numerically investigate the scheme in a more realistic setting where most gates are noisy, including those involved in the parity checks (the only gates that are not noisy are measurements and the circuit that generates the random input state).

Intuitively, given a Clifford circuit and using the \textit{efficient PCS protocol}, we should be able to perform long computations with high fidelity. For a given Clifford circuit, we can keep the $C_2$ checks constant and independent of the depth of $U$. Thus, the noise induced by our $C_2$ checks should be relatively constant. 

The $C_1$ checks depend on $U$, but they are elements of the Pauli group and hence limited in size and complexity. Therefore, the noise induced by the $C_1$ checks should also be limited and independent of the depth of $U$. 

We demonstrate this intuition on simulations consisting of 550 randomly generated Clifford circuits with two compute qubits. Note that these Clifford simulations  provide only intuition that the protocol is suitable for deep circuits because Clifford circuits can in general be optimized to use $O(n^2/log(n))$ \Gate{cnot} gates, where $n$ is the number of qubits \cite{Patel_2008Optimal_Clifford}. Thus, two-qubit Clifford circuits can be optimized to be shallow. It may be possible to prove this performance on Clifford circuits with higher qubit counts. 

\begin{figure}[h]
    \centering
    \includegraphics[scale=0.5]{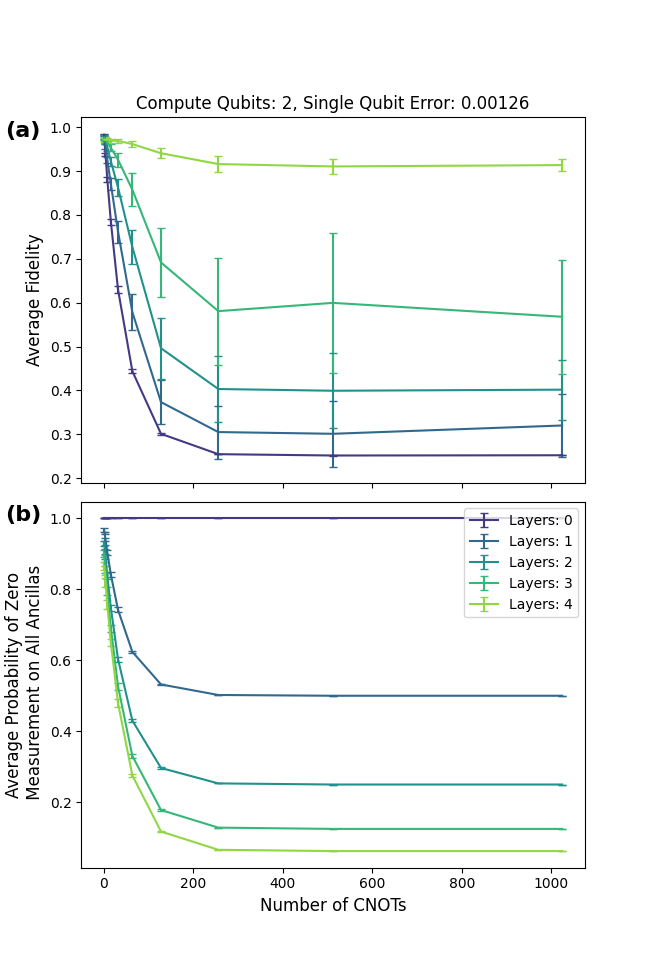}
    \caption{Two-qubit Clifford simulation with single-qubit error of 0.00126 (0.0126 two-qubit error). Note that while these input circuits can be optimized to use $O(n^2/log(n))$ \Gate{cnot} gates (i.e., O(2) \Gate{cnot} gates), these simulations provide intuition that the protocol is suitable for deep circuits.}
    \label{fig:2qubitsLongComputations}
\end{figure}

We considered random circuits with \Gate{cnot} counts that varied from $1, 2, 4, \cdots, 1,024$. For each \Gate{cnot} count we generated $50$ random circuits, and we used single-qubit depolarizing noise of 0.00126 (0.0126 two-qubit noise). This lies within the range of current noise levels found in state of the art quantum computers \cite{maldonado2021_errorReduction, Zhang2020_ErrorMitQuantGatesExceedPhysFidelities}. We used four layers of checks;  the form of the checks was provided in Proposition \ref{lemma:lowNumbOfChecks}. As shown in Fig.~\ref{fig:2qubitsLongComputations}a, we maintained an average fidelity $F_m$ for the postselected state of greater than 90\% for circuits consisting of up to 1,024 \Gate{cnot} gates. The average fidelity of the unmitigated circuits drops to 25\% at 256 \Gate{cnot}s. Note that this comes at the cost of a lower postselection rate of 6.25\% as shown in Fig.~\ref{fig:2qubitsLongComputations}b. 

For optimized Clifford circuits, we would likely not want to use all  the checks from Proposition \ref{lemma:lowNumbOfChecks} because we would probably exceed the \Gate{cnot} count of the input circuit. Still, as shown in Fig.~\ref{fig:2qubitsLongComputations}a, fewer layers can produce significant fidelity improvement.


These simulations establish the general trend that fidelity is positively correlated with the number of layers up to some value. We suspect that these results also hold for general (non-Clifford) circuits. 


We also randomly generated 1,850 input circuits consisting of Clifford + arbitrary diagonal unitary gates. Of these, 1,350 input circuits consist of five qubits with \Gate{cnot} counts of $\{1, 5, \cdots, 40\}$; 500 input circuits consist of ten qubits with \Gate{cnot} counts of $\{1, 5, \cdots, 40, 80\}$. We varied the single-qubit error from $10^{-5}$ to $10^{-2}$ with 21 equally spaced points in log scale.

For the ten-qubit circuits we also generated circuits with \Gate{cnot} gate counts of 80 to match the max \Gate{cnot} count to qubit ratio of the five qubit case. Each random circuit was generated first as a random Clifford gate, which we truncated to reach the desired \Gate{cnot} count. Next, we inserted \Gate{Rz} gates with random rotation angles and random locations in the circuit. We used \Gate{Rz} gate counts of \{5, 10, 15\}. Each \Gate{Rz} value for five qubits consists of 450 circuits. This covers a large class of variational quantum eigensolver  and quantum approximate optimization algorithm  circuits \cite{farhi2014QAOA}. 

\begin{figure}[h]
    \centering
    \includegraphics[scale=0.5]{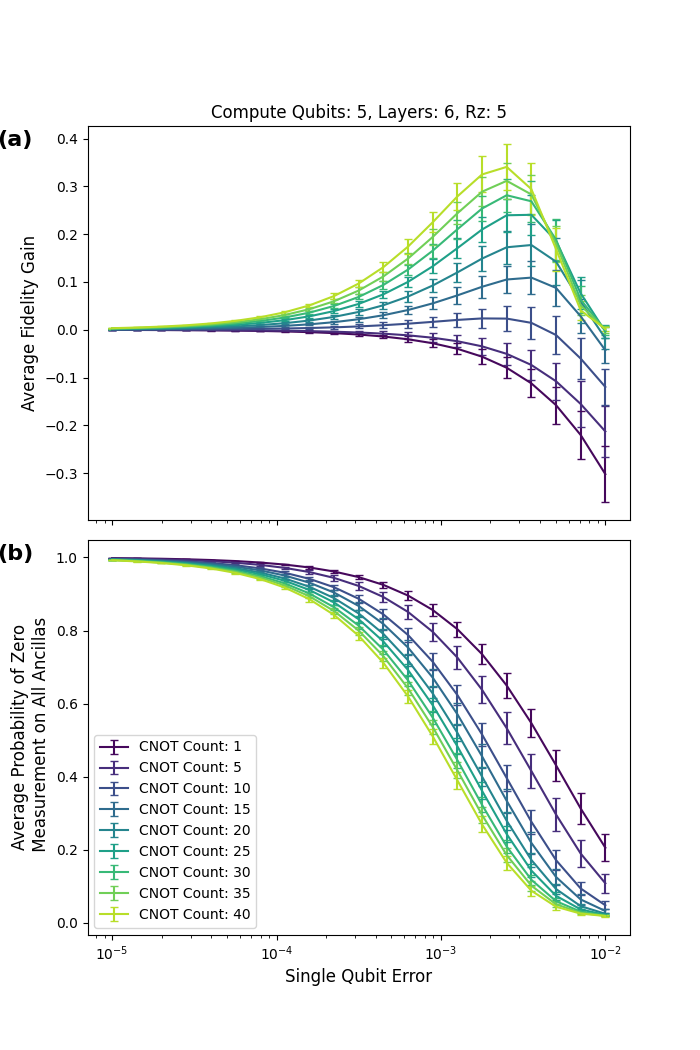}
    \caption{Six layers. (a) The peak average fidelity gain of 34 percentage points occurred at a single-qubit error of approximately 0.00251 (0.0251 two-qubit error) and 40 CNOTs. (b) The probability of postselecting decreases with increasing error rate and increasing \Gate{cnot} count.}
    \label{fig:5qubits_allCNOTs_Layers6_rz5}
\end{figure}
\begin{figure}[h]
    \centering
    \includegraphics[scale=0.5]{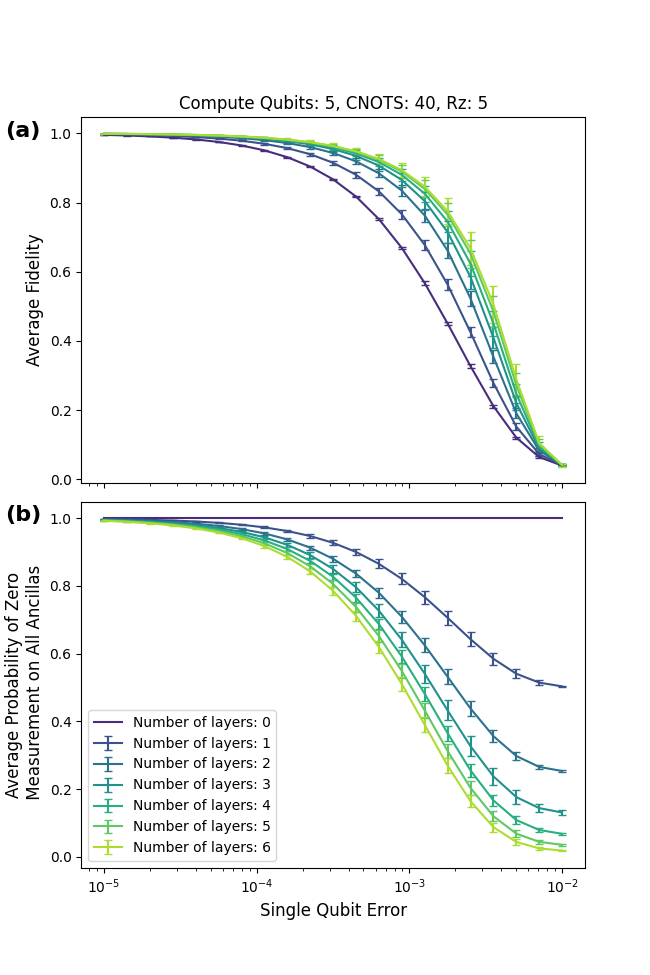}
    \caption{(a) Average fidelity for layers zero to six. At the peak fidelity gain, the nonmitigated circuit has about 33\% fidelity, and the six-layer mitigated circuit has about 67\% fidelity. (b) Probability of postselecting.}
    \label{fig:5qubits_fidelity_40cnots_layers}
\end{figure}

We achieved an average peak fidelity gain $F_m-F_n$  of $34$ percentage points for five-qubit circuits with a \Gate{cnot} gate count of 40, five \Gate{Rz} gates, and six layers of checks, as shown in Fig.~\ref{fig:5qubits_allCNOTs_Layers6_rz5}a. For input circuits with a low \Gate{cnot} count, the fidelity gain is negative because the checks introduce more errors than they eliminate in the post selected state. The average postselection probability is given in Fig.~\ref{fig:5qubits_allCNOTs_Layers6_rz5}b.
We also give  in Fig.~\ref{fig:5qubits_fidelity_40cnots_layers}a  a plot that breaks down this peak fidelity gain. 
At the peak fidelity gain, the nonmitigated circuit has about 33\% fidelity, and the six-layer mitigated circuit has about 67\% fidelity.
As shown in Fig.~\ref{fig:5qubits_fidelity_40cnots_layers}a, the mitigated circuits perform significantly and consistently better than the unmitigated circuits. Even for lower-layer counts such as two, the average fidelity gain reached 20 percentage points. Fig.~\ref{fig:5qubits_fidelity_40cnots_layers}b gives the corresponding post selection probabilities and demonstrates that we have significant control over the probabilities by changing the number of layers.

\begin{figure}[h]
    \centering
    \includegraphics[scale=0.5]{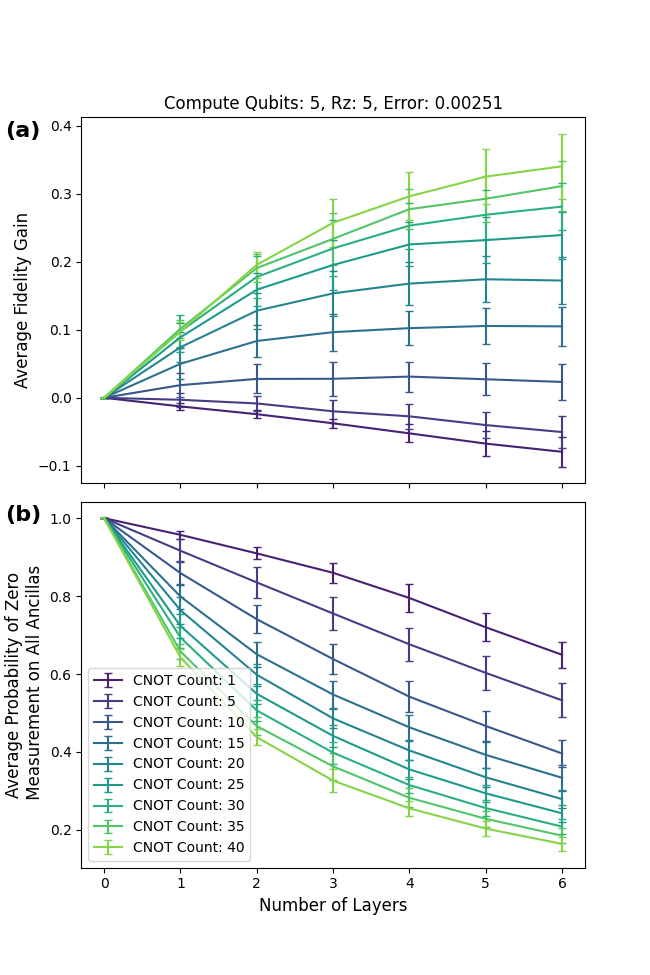}
    \caption{(a) Average fidelity gain vs number of layers. (b) Probability of postselecting vs number of layers. The single-qubit error is fixed at approximately 0.00251 (0.0251 two-qubit error)}
    \label{fig:5qubits_fidelity_vs_layers_rz5}
\end{figure}

Each additional layer increased the average fidelity provided enough circuit depth. We show this in more detail in Fig.~\ref{fig:5qubits_fidelity_vs_layers_rz5}a, where we fixed the single-qubit error rate to 0.00251 (0.0251 two-qubit error) the value that gave the peak fidelity gain in Fig.~\ref{fig:5qubits_allCNOTs_Layers6_rz5}a.
Circuits with more than six layers  may result in even better performance, but the amount of fidelity gained decreases with subsequent layers. Fig.~\ref{fig:5qubits_fidelity_vs_layers_rz5}b shows the corresponding post selection probabilities and the minimum post selection probability is about 16\%.

\begin{figure}[h]
    \centering
    \includegraphics[scale=0.5]{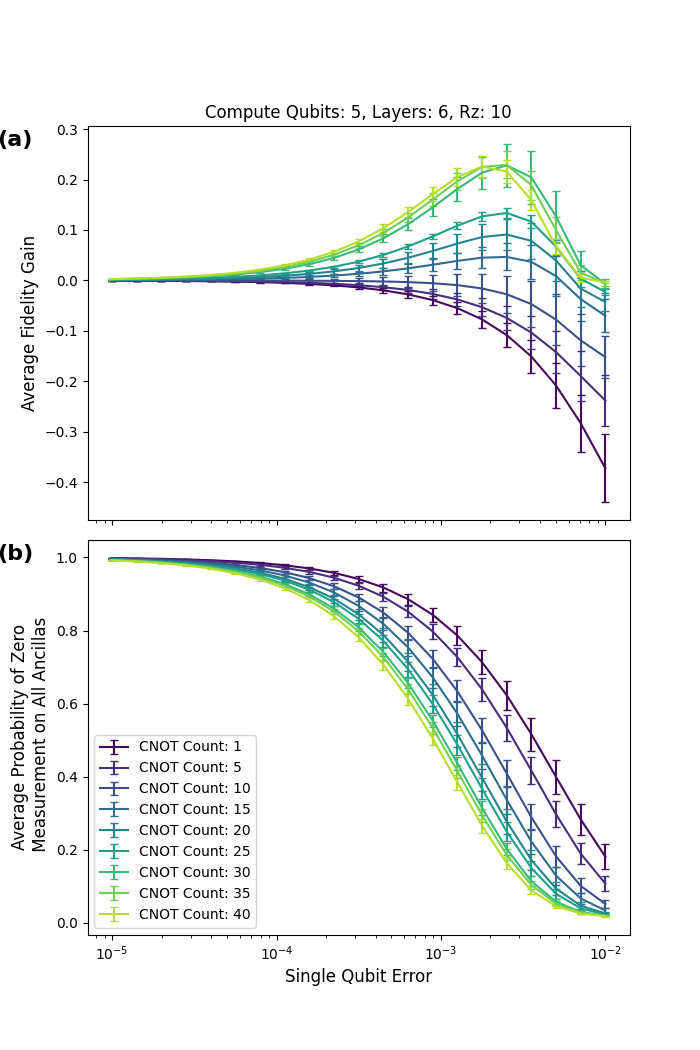}
    \caption{Five-qubit circuits with 10 \Gate{Rz} gates. (a) The max fidelity gain is about 25 percentage points. (b) Probability of postselecting vs single-qubit error.}
    \label{fig:5qubits_fidelity_rz10_layers_6}
\end{figure}

\begin{figure}[h]
    \centering
    \includegraphics[scale=0.5]{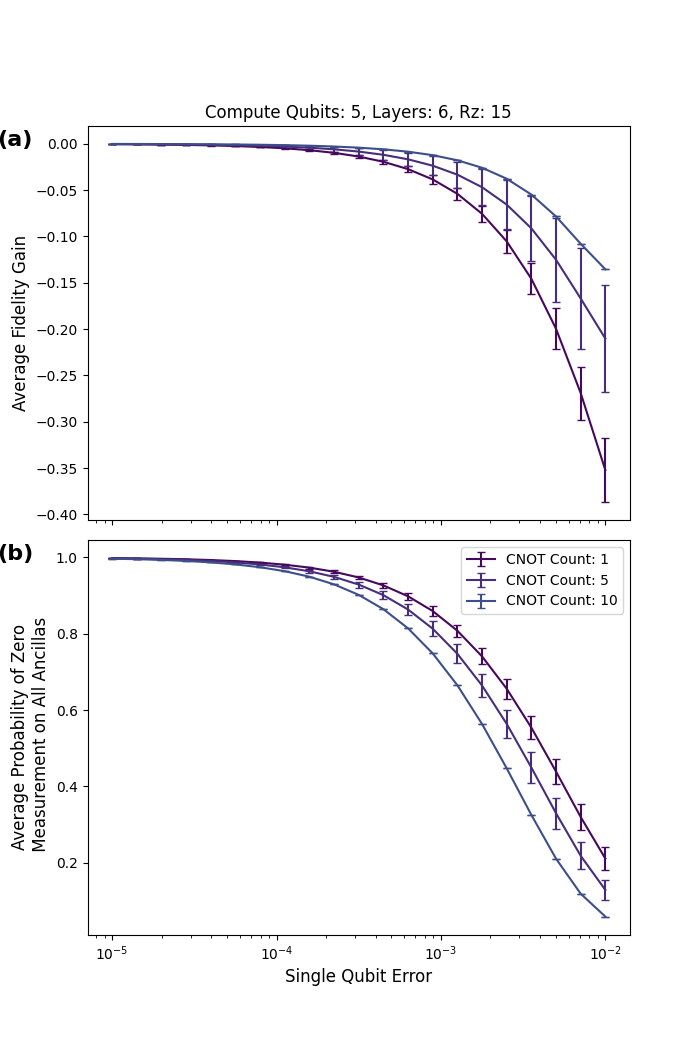}
    \caption{Five-qubit circuits with 15 \Gate{Rz} gates. After 10 \Gate{cnot} gates, we cannot find circuits with six layers.}
    \label{fig:5qubits_fidelity_rz15_layers_6}
\end{figure}

\begin{figure}[h]
    \centering
    \includegraphics[scale=0.5]{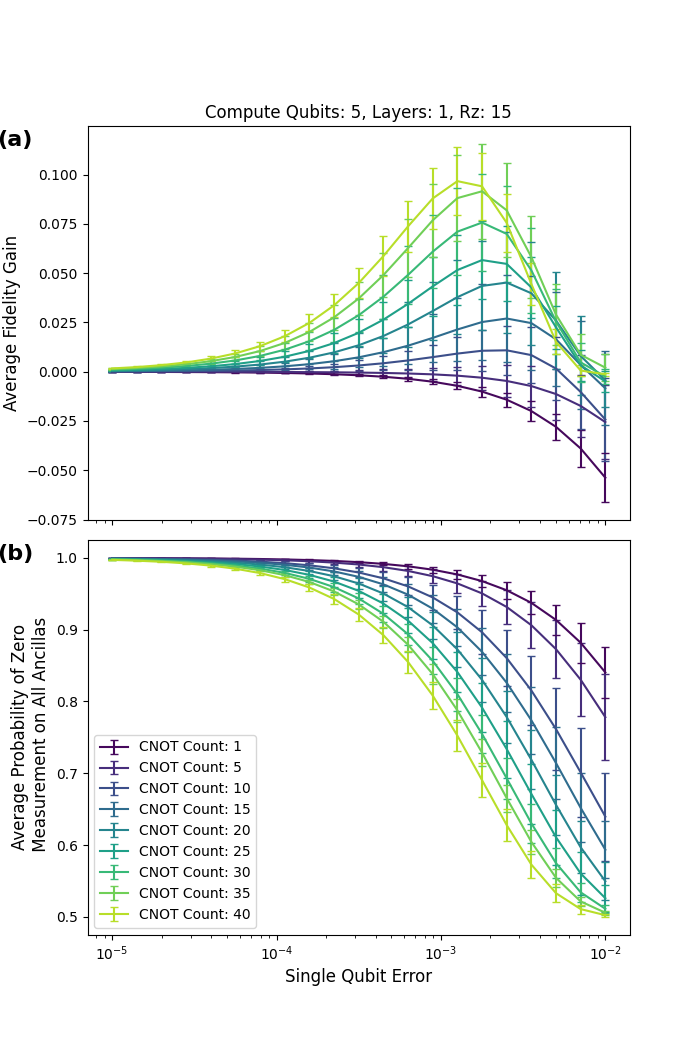}
    \caption{Five-qubit circuits with 15 \Gate{Rz} gates. (a) At one layer of checks, the peak fidelity gain is about 10 percentage points.}
    \label{fig:5qubits_fidelity_rz15_layers_1}
\end{figure}

As the number of \Gate{Rz} (non-Clifford) gates increases, the number of possible low weight $\tilde C_2$ checks for the \textit{efficient PCS protocol} decreases, and consequently the fidelity gain decreases. As shown in Fig.~\ref{fig:5qubits_fidelity_rz10_layers_6}a, at an \Gate{Rz} gate count of 10, the peak fidelity gain is about 25\%. 
As shown in Fig.~\ref{fig:5qubits_fidelity_rz15_layers_6}a, at an \Gate{Rz} gate count of 15, we cannot find six layers of checks for random circuits with 20 \Gate{cnot} gates or higher. Interestingly, as shown in Figs.~\ref{fig:5qubits_allCNOTs_Layers6_rz5}b, \ref{fig:5qubits_fidelity_rz10_layers_6}b and \ref{fig:5qubits_fidelity_rz15_layers_6}b, the post selection curves are relatively unchanged.
Using one layer of checks, we have a peak fidelity gain of about 10 percentage points at 40 \Gate{cnot} gates, as shown in Fig.~\ref{fig:5qubits_fidelity_rz15_layers_1}a. Fig.~\ref{fig:5qubits_fidelity_rz15_layers_1}b shows the corresponding post selection probabilities.

\begin{figure}[h]
    \centering
    \includegraphics[scale=0.5]{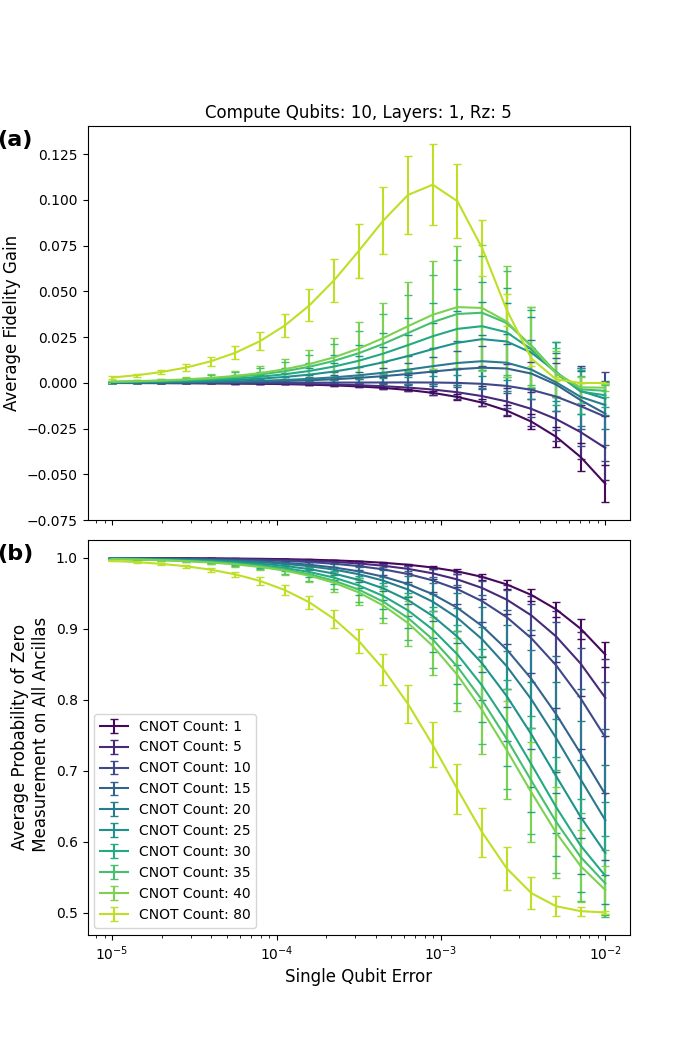}
    \caption{Ten-qubit circuits with 5 \Gate{Rz} gates. We used a single layer of low weight checks. The 80 \Gate{cnot} count case matches the \Gate{cnot} count to qubit ratio of the five-qubit case with 40 \Gate{cnot} gates. (a) The peak fidelity gain is about ten percentage points. It occurs at a single-qubit error of about 0.000891 (0.00891 two-qubit error)}
    \label{fig:10qubits_low_weight}
\end{figure}

For the ten-qubit case, as shown in Fig.~\ref{fig:10qubits_low_weight}a, we achieved a fidelity gain of about ten percentage points. This occurred at a \Gate{cnot} count to qubit ratio of eight, which matches the scenario of the peak fidelity gain in the five-qubit case. The peak fidelity gain occurred at a single-qubit error of about 0.000891 (0.00891 two-qubit error). Fig.~\ref{fig:10qubits_low_weight}b shows the corresponding post selection probabilities.

\begin{figure}[!h]
    \begin{subfigure}{.5\textwidth}
        \centering
        \includegraphics[scale=0.5]{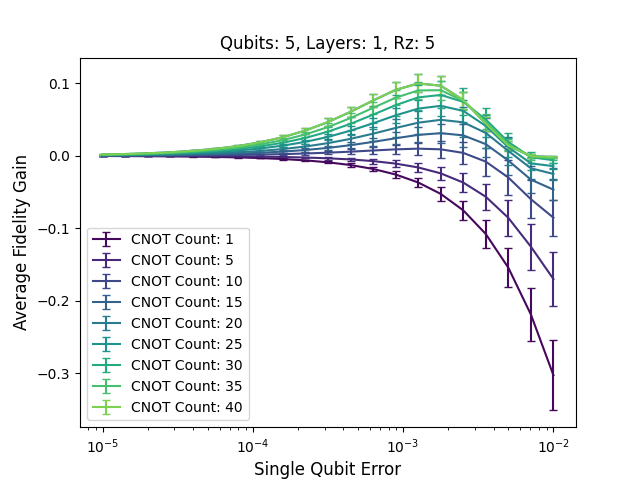}
        \caption{One layer using max weight checks for five compute qubits.}
        \label{fig:5qubits_all}
    \end{subfigure}
    \begin{subfigure}{.5\textwidth}
        \centering
        \includegraphics[scale=0.5]{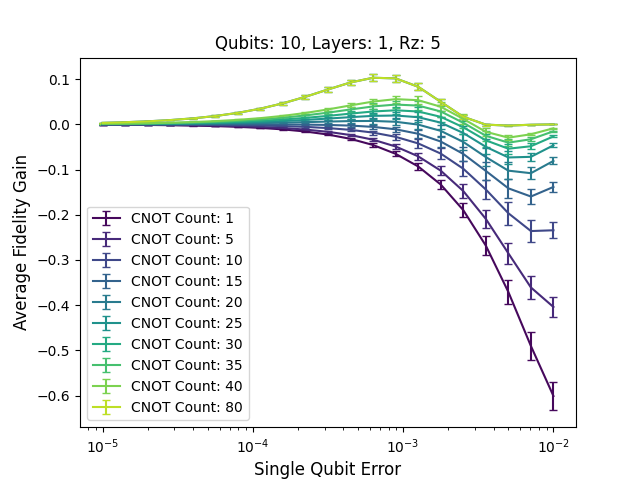}
        \caption{One layer using max weight checks for ten compute qubits.}
        \label{fig:10qubits_all}
    \end{subfigure}
    \caption{Max weight checks. The high-weight checks introduce a lot of noise compared with the low-weight method, as shown in the large negative fidelity at a high single-qubit error rate.}
\end{figure}

The preceding simulations focus on using low-weight checks first. We now analyze the performance of high-weight checks. As shown in Figs.~\ref{fig:5qubits_all} and \ref{fig:10qubits_all}, while the high-weight checks do give a boost in fidelity, they introduce significant amounts of noise compared to the low-weight checks.

\section{Conclusions}\label{subsec:conclusions}
The quantum error mitigation technique we have studied in this  work is novel because (1) it has an adjustable quantum overhead for any input circuit, (2) by adjusting the number of layers of check operators, the technique allows controlling of the post-selection probability and the error from the error mitigation protocol, (3) the method can be applied repeatedly and at any location in the circuit, and works for arbitrary input states, and (4) in the setting of Theorem \ref{thm:MultilayerUnitFidelity}, we prove that we can achieve unit fidelity provided that we use a sufficient number of layers.

We prove in Theorem \ref{thm:MultilayerUnitFidelity} that if the error is restricted to the compute qubits (see Fig.~\ref{fig:multilayerNoisy}), there exist checks such that the fidelity for the postselected state reaches unity. We also give a small number of $\tilde C_2$ checks that reach unit fidelity in this scenario in Propositions \ref{lemma:weightOneErrors} and \ref{lemma:lowNumbOfChecks}.


In Eq.~\eqref{eq:multiLayerConditionOnChecks1}, $\tilde C_2$ is chosen and $\tilde C_1$ can be directly determined through our \textit{finding checks protocol} given in Section \ref{subsec:FindingChecks}. This algorithm determines the pairs of checks without matrix multiplication. Instead, we perform lookups of predetermined commutation relations. One limitation of our \textit{finding checks protocol} is that we are able to find only $\tilde{C}_1$ that are in the Pauli group. This limitation does not exist for the \textit{general PCS protocol}.

The main limitation of the proposed approach is the need to obtain the checks $\tilde{C}_1$ and $\tilde{C}_2$, with cost exponential in the number of qubits in the subcircuit. This cost can be reduced to exponential in the number of non-Clifford gates (and only polynomial in the number of qubits) by leveraging the extended stabilizer formalism~\cite{Bravyi2019}.


The performance of the protocol is tested through extensive numerical simulations on random circuits consisting of 550 Clifford and 1,850 non-Clifford circuits. We used the Clifford simulations to provide intuition that the technique is suitable for deep circuits. 

For the non-Clifford circuits, we used five- and ten-qubit circuits. We use the difference between the fidelity of the mitigated circuit and the fidelity of the unmitigated circuit as a figure of merit. Under depolarizing noise, the simulations reached an average fidelity gain of $34$ percentage points for circuits consisting of five qubits, 40 CNOTs, and six low-weight $\tilde{C}_2$ checks (see Figs.~\ref{fig:5qubits_allCNOTs_Layers6_rz5}a and \ref{fig:5qubits_allCNOTs_Layers6_rz5}b). It is possible that more layers will provide further boosts in fidelity. The single-qubit noise ranged from $10^{-5}$ to $10^{-1}$. This coincides with current noise levels found in superconducting quantum computers \cite{maldonado2021_errorReduction}.

In \cite{Xiong_2021CircSymmVerifMitigQuantDomImpair}, the authors derive an error mitigation scheme based on symmetry verification, which they call the spatio-temporal
stabilizer (STS) technique. The STS technique shares many similarities with the PCS scheme, as first introduced in \cite{Debroy_2020ExtendedFlagGadgets}, and when there is only one pair of checks, STS is the PCS scheme. An important difference is that when there are multiple pairs of checks, 
layers are allowed to be partly nested in the STS technique. For example, a possible STS execution is layer one and layer two act on the same compute qubits, but layer two begins before layer one has ended and layer one ends before layer two. Since the STS method also allows the standard layering of checks in PCS, our results also hold for the STS technique.

We also note that while the results of this research are presented in the context of quantum computing, the theoretical results hold in general for settings where the user intends to implement an ideal known unitary $U$ on a quantum state. This follows because we placed no restrictions on the unitary. The performance of the scheme in other settings needs to be investigated. Also, since the protocol places no restriction on the input state, one can apply the mitigation technique on subcircuits and easily combine it with other methods. Splitting a large circuit into subcircuits for finding checks or combining the protocol with other techniques have not been studied. Determining the optimal number of check layers also needs to be further investigated. 

Moreover, the best type of checks to use may be non-Pauli in the \textit{general PCS protocol}. This is likely true given some knowledge of the dominant noise. One potential line of investigation is to use the controls in dynamical decoupling protocols as the $\tilde{C}_2$ parity checks \cite{Khodjasteh_2007DD, Tripathi_2021SuppOfCrosstalkDynamicalDecoupling}.

\section{Data Availability}
The data presented in this paper is available online at
\url{https://github.com/alvinquantum/noise_mitigation_symmetry}.

\section{Code Availability}
The code used for numerical experiments in this work is available online at \url{https://github.com/alvinquantum/noise_mitigation_symmetry}.

\section*{Acknowledgments}
This research was supported in part by an appointment to the Intelligence Community Postdoctoral Research Fellowship Program at Argonne National Laboratory, administered by Oak Ridge Institute for Science and Education through an interagency agreement between the U.S. Department of Energy and the Office of the Director of National Intelligence. This work was supported in part by the U.S.\ Department of Energy (DOE), Office of Science, National Quantum Information Science Research Centers, Office of Advanced Scientific Computing Research AIDE-QC and FAR-QC projects, and by the Argonne LDRD program under contract number DE-AC02-06CH11357. Z.S.  thanks Q-NEXT  for supporting this project. We gratefully acknowledge the computing resources provided on Bebop, a high-performance computing cluster operated by the Laboratory Computing Resource Center at Argonne National Laboratory.

\section*{Author Contributions}
AG performed the theoretical analysis with feedback from RS and ZS. RS developed the ``finding checks quickly" protocol. ZS suggested using the Pauli expansion of Kraus operators in the theory.  AG wrote the code, performed the numerical simulations, and wrote the manuscript with input from RS, ZS, and MS. RS structured multiple sections of the manuscript. RS, ZS, and MS developed the numerical experiments with input from AG. RS, MS, AG, and ZS found numerical optimizations. 

\section*{Competing Interests}
The authors declare no competing interests.

\section*{Disclaimer}

This paper was prepared for information purposes with contributions from the Global Technology Applied Research group of JPMorgan Chase. This paper is not a product of the Research Department of JPMorgan Chase. or its affiliates. Neither JPMorgan Chase nor any of its affiliates make any explicit or implied representation or warranty and none of them accept any liability in connection with this paper, including, but not limited to, the completeness, accuracy, reliability of information contained herein and the potential legal, compliance, tax or accounting effects thereof. This document is not intended as investment research or investment advice, or a recommendation, offer or solicitation for the purchase or sale of any security, financial instrument, financial product or service, or to be used in any way for evaluating the merits of participating in any transaction.

\bibliographystyle{unsrt}


\vfill

\small

\framebox{\parbox{\linewidth}{
The submitted manuscript has been created by UChicago Argonne, LLC, Operator of 
Argonne National Laboratory (``Argonne''). Argonne, a U.S.\ Department of 
Energy Office of Science laboratory, is operated under Contract No.\ 
DE-AC02-06CH11357. 
The U.S.\ Government retains for itself, and others acting on its behalf, a 
paid-up nonexclusive, irrevocable worldwide license in said article to 
reproduce, prepare derivative works, distribute copies to the public, and 
perform publicly and display publicly, by or on behalf of the Government.  The 
Department of Energy will provide public access to these results of federally 
sponsored research in accordance with the DOE Public Access Plan. 
http://energy.gov/downloads/doe-public-access-plan.}}

\end{document}